\input mn.tex
\input psfig.tex

\pageoffset{-2.5pc}{0pc}


\pagerange{000-000}    
\pubyear{1999}
\volume{000}

\begintopmatter  

\title{Radiation force on relativistic jets in active galactic nuclei}

\author{Qinghuan Luo and R. J. Protheroe}

\affiliation{Department of Physics and Mathematical Physics,
The University of Adelaide, Adelaide, SA 5005, Australia}

\acceptedline{Accepted date; Received date}

\shortauthor{Luo and Protheroe}

\shorttitle{Relativistic jets}

\abstract{
Radiative deceleration of relativistic jets in active galactic nuclei 
as the result of inverse Compton scattering of soft photons from 
accretion discs is discussed. The Klein-Nishina (KN) cross section is used
in the calculation of the radiation force due to inverse Compton 
scattering.  Our result shows that deceleration due to scattering in 
the KN regime is important only for jets starting with a bulk Lorentz 
factor larger than $10^3$. When the bulk Lorentz factor satisfies this
condition, particles scattering in the Thomson regime contribute
positively to the radiation force (acceleration), but those particles
scattering in the KN regime are dominant and the overall effect is 
deceleration. In the KN limit, the drag due to Compton 
scattering, though less severe than in the Thomson limit, strongly 
constrains the bulk Lorentz factor. Most of the power from the deceleration
goes into radiation and hence the ability of the jet to transport
significant power (in particle kinetic energy) out of the subparsec 
region is severely limited. The deceleration efficiency
decreases significantly if the jet contains protons and the proton
to electron number  density ratio satisfies the condition 
$n_p/n_{e0}>2\gamma_{\rm min}/\mu_p$ where $\gamma_{\rm min}$ is the 
minimum Lorentz factor of relativistic electrons (or positrons) in the 
jet frame and $\mu_p$ is the proton to electron mass ratio.  
}

\maketitle

\section{Introduction}

High energy observations of blazars strongly suggest that the emission
comes from relativistic jets of active galactic nuclei (AGN) (e.g. 
von Montigny et al. 1995; Thompson et al. 1995; Gaidos  et al. 1996;
Quinn et al. 1996; Schubnell et al. 1996). Other evidence for relativistic
AGN jets includes the observation of variability in radio, optical, and 
high energy gamma ray emission from blazars.  Although there is no 
consensus on the details of AGN, it is widely accepted that a jet may 
form near a black hole with an accretion disc (e.g. a review by 
Blandford 1990). Since a substantial fraction of the binding energy 
of accreting material is dissipated and converted to 
radiation, the disc is a strong source of soft photons (e.g. as black 
body radiation or reprocessed radiation). Interaction of relativistic 
particles in the jet with these surrounding photon fields can be 
important, and may contribute to the observed gamma ray emission, e.g.
through inverse Compton scattering (Reynolds 1982;
Melia \& K\"onigl 1989; Dermer \& Schlickeiser 1993). Disk emission 
can also affect the jet dynamics near the black hole through radiative 
acceleration or deceleration.  The effect of Compton scattering on 
the jet flow was discussed by O'Dell (1981), and Phinney (1982, 1987), 
and it was suggested that under certain (quite stringent) conditions, the 
jet can be accelerated through Compton scattering. However, as pointed 
out by several authors (e.g. Phinney 1987; Melia \& K\"onigl 1989; Sikora 
et al. 1996), in the region close to the black hole, Compton scattering 
can be more effective in slowing down rather than accelerating the jet,
and can constrain the bulk flow of the jet in that region.

There are AGN emission models in which protons are accelerated to ultra 
high energies (e.g. K\"onigl 1994). One of the attractive features of this 
type of model is that protons can be accelerated to high energies 
without significant energy loss. These high energy protons can interact
with photon fields producing electron-positron pairs. It is possible that
through pair cascades, an ultrarelativistic jet is produced. Alternatively,
an ultrarelativistic jet can be produced through rapid acceleration 
of $e^\pm$ such as by magnetic reconnection (e.g. Haswell et al. 1992)
or by rotation-induced electrostatic potential drop as in pulsars 
(Michel 1987). In both cases, the bulk flow of the jet may initially 
have a large Lorentz factor, and plasmas in the jet can be highly 
relativistic.  Therefore, scattering in the Klein-Nishina (KN) regime 
is important, and should be included in calculation of radiation 
force due to inverse Compton scattering. There is then the question of 
whether the jet with initially large bulk flow speed is subject to the same 
(severe) radiation deceleration as in the Thomson regime, and this will 
be explored in this paper.

The constraint on the bulk flow by Compton scattering should strongly
depend on the soft photon distribution of the disc. In our discussion,
disc emission is assumed to be axisymmetric, and so the only effect 
on the jet flow considered is in the jet direction (it is usually 
assumed that the jet is normal to the disc plane). However, there are 
cases in which accretion may not be circular, e.g. eccentric 
accretion can occur if the black hole is in a binary system (e.g.
Begelman, Blandford \& Rees 1980; Eracleous et al. 1995), and this 
possibility will be considered in a forthcoming paper (Luo 1998).

The deceleration efficiency strongly depends on the composition of the
jet. An AGN jet that contains protons would be subject to less severe
deceleration because protons scatter photons with much smaller cross
section than that for electrons (or positrons) and this effectively
increases the jet inertia.

In this paper, we extend calculations of the radiation force by 
including the Klein-Nishina effect as the earlier calculations were 
done only in the Thomson scattering regime (e.g. O'Dell 1981; Phinney 
1982; Sikora et al. 1996). In Sec. 2, the average radiation 
force on the bulk plasma flow is derived in both the Thomson and 
Klein-Nishina limits. Constraints by radiative deceleration on the 
Lorentz factor of the jet are discussed in Sec. 3.

\section{Radiative effect on jet flow}

Through Compton scattering of external photons, individual particles 
in plasmas lose energy and at the same time there is momemtum 
transfer to the plasma. For plasmas in a jet,  the 
momemtum transfer can affect the jet dynamics, i.e. the bulk flow 
can be either accelerated  or decelerated (i.e. radiative drag). 

Consider a comoving cell with energy $\tilde{E}$ (in $m_ec^2$ per
unit volume) in a relativistic jet with the bulk Lorentz factor 
$\Gamma=1/(1-\beta^2_{_{\rm b}})^{1/2}$ where $\beta_{_{\rm b}}$ is 
the bulk velocity in $c$. The tilded quantities correspond to the values
in the lab frame ($K$). We assume that the jet contains mainly 
electron-positron pairs (there may be protons as well but in small
number) and that within the cell electron-positron pair production is not 
important. The latter assumption may not be accurate since pairs
can be injected continuously along the jet. Nonetheless, it allows
us to single out the radiative effect due to Compton scattering, 
and the result should not change qualitatively when the pair production
effect is included. Let $f$ be the radiative force (in $m_ec$ per second
per unit volume) on the cell seen in the jet comoving frame ($K_j$). 
Then, the rate
of change of $\Gamma$ due to radiation force on the  cell is given 
by (e.g.  Phinney 1982; Sikora et al. 1996) 
$${d\Gamma\over dt}={\beta_{_{\rm b}}f_z\over E}, \eqno(2.1)$$
where $E$ is the cell energy in $K_j$, and  we assume that the jet is
along the $z$ direction. Then, for $f_z>0$, the jet is accelerated away from 
the source, and for $f_z<0$, it is decelerated toward the source, i.e. 
radiation drag. In (2.1), the average force $f_z$ depends on the particle 
distribution which is influenced by the specific acceleration or 
energy loss mechanisms. In the following discussion, we only consider
radiative effects due to Compton scattering.

\subsection{Radiation from the disc}

To study the radiative effect on jet dynamics due to inverse Compton 
scattering, a specific model for the external photon field is required.
We assume that the external photon field is the accretion disc emission, 
and is axisymmetric with respect to the jet axis (which is in the $z$ 
direction). A schematic diagram for such a disc-jet system is shown in 
Figure 1. Emission from the disc is modeled as the sum of emission from 
series of rings centered at the black hole; each of them emits photons 
with a characteristic energy $\tilde{\varepsilon}_{_R}$ where $R$ is the
radius of the ring. The flux number 
density of photons from the ring $R\sim R+dR$ is given by
$F(\tilde{\varepsilon},R)=F(R)\delta(\tilde{\varepsilon}
-\tilde{\varepsilon}_{_R})$ where $\tilde{\varepsilon}
=\Gamma(1+\beta{_{\rm b}}\,\cos\theta)\varepsilon$ is the photon energy
(in $m_ec^2$) seen in $K$, $\varepsilon$ is the corresponding energy
seen in $K_j$, $\theta$ is the angle between the photon propagation
direction and the jet axis, and  $F(R)$ is given by (Blandford 1990)
$$\eqalign{F(R)\approx&4.4\times10^{30}\biggl({1\,{\rm
eV}\over\tilde{\varepsilon}_{_R}} \biggr)
\biggl({\dot{M}\over L_{\rm Edd}/c^2}\biggr)\cr
&\times \biggl({M\over 10^8M_\odot}\biggr)^{-1}\biggl(
{R\over GM/c^2}\biggr)^{-3}I\,{\rm cm}^{-2}\,{\rm s}^{-1},\cr}\eqno(2.2)$$
where $I=(1-R_{\rm min}/R)^{1/2}$, $R_{\rm min}$ is the radius of the 
innermost part of the disc, $M$ is the mass of the black hole, $G$ 
is Newton's constant, $L_{\rm Edd}$ is the Eddington luminosity. We 
assume $R_{\rm max}$ to be the maximum radius of the disc, within 
which emission can be adequately described by (2.2). 

A detailed model for emission from the accretion disc should include 
effects such as reprocessing of radiation, disc corona, etc. As we are
concerned with the radiative effect on the jet flow, all these details
will be ignored and we assume that disc emission is blackbody with 
$\tilde{\varepsilon}_{_R} \approx 2.7k_{\rm B}T(R)/m_ec^2$, where $T(R)$ 
is the effective temperature of emission from the ring with $R$. Then, 
for an optically thick disc, $T(R)$ is identified as the surface temperature,
and it is given by $T(R)=(3GM\dot{M}I/8\pi R^3\sigma_{\rm SB})^{1/4}$,
that is,
$$\eqalign{T(R)\approx&5\times10^5\!
\biggl(\!{\dot{M}\over L_{\rm Edd}/c^2}\!\biggr)^{\!1/4}
\biggl(\!{M\over10^8M_\odot}\!\biggr)^{\!-1/4}\cr
&\times
\biggl(\!{R\over GM/c^2}\!\biggr)^{\!-3/4}I^{1/4}\,{\rm K},\cr}\eqno(2.3)$$
where $R\geq R_{\rm min}$, $\sigma_{\rm SB}$ is the Stefan-Boltzmann 
constant. Figure 2 shows plots of $F(R)$ for $R_{\rm min}/R_g=5$ and 6, 
where $R_g=GM/c^2\approx1.48\times10^{13}\,{\rm cm}\,(M/10^8\,M_\odot)$
is the gravitational radius. The fluxes are peaked at 
$R_0=(7/6)R_{_{\rm min}}\approx 5.8R_g$ and $7R_g$, respectively.
 
Let $n_{\rm ph}(\varepsilon,\Omega,R)dR$ be the number density 
of photons emitted from the ring between $R$ and $R+dR$, where 
$\varepsilon$ and $\Omega=(\phi,\cos\theta)$ are the energy and direction
of incoming photons, respectively. Using (2.2), the photon number 
density (per unit $R$), $n_{\rm ph}(\varepsilon,\Omega,R)$, in $K_j$ 
can  be written as (e.g. Dermer \& Schlickeiser 1993) 
$$n_{\rm ph}(\varepsilon,\Omega,R)={RF(\tilde{\varepsilon},R)\over 2\pi
c(R^2+z^2)}\,\delta\biggl( \cos\theta-{\cos\tilde{\theta}_{_R}-\beta_{_{\rm
b}}\over 1-\beta{_{\rm b}}\,\cos\tilde{\theta}_{_R}}\biggr),\eqno(2.4)$$
where $\cos\tilde{\theta}_{_R}=z/(R^2+z^2)^{1/2}$, $z$ is the distance
from the disc surface, $\tilde{\theta}_{_R}$ is the angle of incoming 
photons relative to the the jet axis in $K_j$. 

\beginfigure{1}
\psfig{file=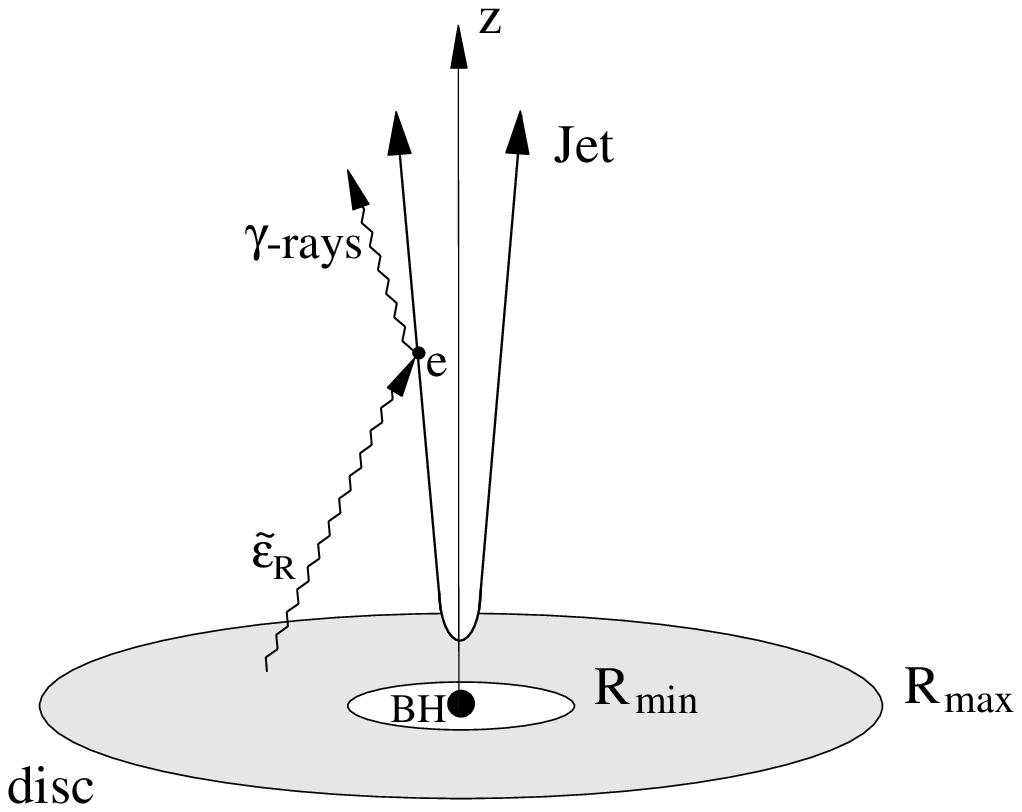,height=6cm,width=7cm}
\medskip
\caption{{\bf Figure 1.} A schematic diagram of an AGN jet.}
\endfigure

\beginfigure{2}
\psfig{file=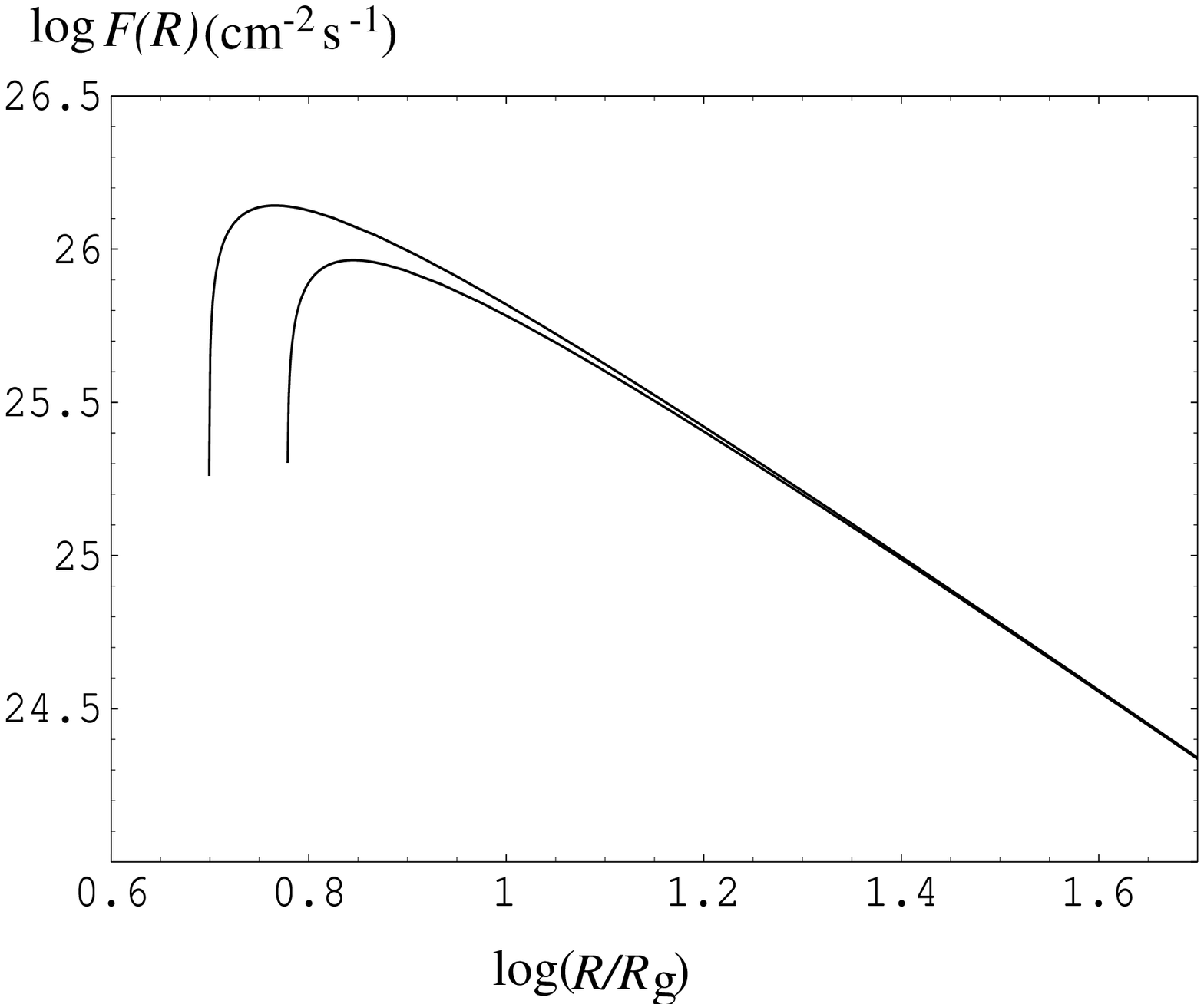,height=7cm,width=8cm}
\caption{{\bf Figure 2.} Photon flux density from the ring  as a function 
of $R$. We assume $M=10^8M_\odot$, $\dot{M}=0.1L_{\rm Edd}/c^2$.
The upper and lower curves are obtained by assuming $R_{\rm min}=5R_g$ and
$6R_g$, respectively.}
\endfigure

\subsection{Radiation force}

The radiation force $f$ (on the jet bulk flow)  can be derived by 
calculating the rate of average momemtum transfer to plasmas in
the jet as a result of inverse Compton scattering, which is given by
$$\eqalign{{\bf f}&=\int\!d\Omega_ed\gamma\,{d{\bf p}\over dt}\,
n_e(\gamma,\Omega_e)\cr
=&-c\!\int\!dR\int\!d\Omega_ed\gamma\,n_e(\gamma,\Omega_e)\!
 \int\!\!d\varepsilon\!\int\!\!d\Omega\,
n_{_{\rm ph}}(\varepsilon,\Omega,R)\,D\cr
\times&\int\!\!d\varepsilon_s
\int\!\! d\Omega_s\Biggl({d\sigma\over d\varepsilon_s
d\Omega_s}\biggr)(\varepsilon_s\hat{\bf k}_s-
\varepsilon\hat{\bf k}),\cr}
\eqno(2.5)$$
where $D=1-{\bf \beta}\cdot\hat{\bf k}$, $\gamma=1/(1-\beta^2)^{1/2}$
is the particle Lorentz factor in the jet frame ($K_j$),
$n_{_{\rm ph}}(\varepsilon,\Omega)$ is the photon density,
${\bf p}=\gamma{\bf \beta}$ is the momemtum (in $m_ec$),
$n_e(\gamma, \Omega_e)$ is the particle distribution,
$\Omega_e=(\phi_e,\cos\theta_e)$ is the direction of electron
motion, $d\sigma/d\varepsilon_sd\Omega_s$ is the differential cross in $K_j$
whose approximations in the Thomson and KN regimes are given respectively
by (A1) and (A2). The direction of incoming and scattered photons in 
$K_j$ are represented by $\hat{\bf k}$ and $\hat{\bf k}_s$, 
respectively. All quantities with subscript $s$ are for scattered photons. 
As we use axisymmetric disk emission as given by (2.2), the only 
relevant component of the force is along $z$ axis. 

We first derive the radiation force in the Thomson limit.  In the 
Thomson scattering regime, scattering is elastic and we have 
$\varepsilon'_s\approx\varepsilon'$ where $\varepsilon'$ and
$\varepsilon'_s$ are respectively the energies of the incoming and 
scattered photon in the electron rest frame ($K'$). Further, we assume 
a beaming approximation in which the scattered photons propagate 
approximately in the electron's direction in $K_j$ provided that 
$\gamma\gg1$. Based on these considerations, an approximation for the 
differential cross section in $K_j$ can be obtained as (A1), and the 
force is calculated as   
$$\eqalign{f_z=&\sigma_{_{\rm T}}\!\int^{R_{\rm
max}}_{R_{\rm min}}\!dR\, {R\,F(R)\over
R^2+z^2}\,\Gamma^2\tilde{\varepsilon}_R
(1-\beta_{_b}\cos\tilde{\theta}_{_R})\cr
&\times(\cos\tilde{\theta}_{_R}-\beta_{_b})
\Bigl(\textstyle{2\over3}\langle\gamma^2\beta\rangle+1\Bigr)n_{e0},\cr}
\eqno(2.6)$$ 
where we assume that the disc emission is important within
$R_{\rm min} \leq R\leq R_{\rm max}$, $\sigma_{_T}$ is the Thomson cross
section, the average is made over the electron (or positron) distribution,
$\langle...\rangle=\int d\Omega_ed\gamma\,(...)n_e(\gamma, \Omega_e)/n_{e0}$,
and $n_{e0}$ is the average number density. The distribution 
$n_e(\gamma, \Omega_e)$ is assumed to be isotropic in $K_j$ with a 
power law:
$$n_e(\gamma,\Omega_e)={C_0\over4\pi}n_{e0}\gamma^{-p},\eqno(2.7)$$
where $p$ is the electron spectral index with values between 2 and 3,
$\gamma_{\rm min}\leq\gamma\leq\gamma_{\rm max}$, the constant $C_0$ 
is chosen such that (2.7) is normalized to $n_{e0}$. We assume the 
cyclotron time scale is much less than the relevant time scales of 
acceleration or deceleration, electron (positron) energy loss. 
Therefore, all quantities in (2.7) can be regarded as average over
the cyclotron time. For 
$\gamma_{_{\rm max}} \gg\gamma_{_{\rm min}}$ and $p\geq2$, we have 
$C_0\approx (p-1)\gamma^{p-1}_{_{\rm min}}$.
Using (2.7), we have $\langle\gamma^2\rangle\approx[(p-1)/(3-p)]
\gamma^{p-1}_{_{\rm min}}(\gamma^{3-p}_{_{\rm max}}-
\gamma^{3-p}_{_{\rm min}})$ which reduces to 
$\langle\gamma^2\rangle\approx 2\gamma^2_{_{\rm min}}
\ln(\gamma_{_{\rm max}}/\gamma_{_{\rm min}})$ for $p=3$.

It follows from (2.6) that acceleration ($f_z>0$) occurs for 
$\cos\tilde{\theta}_{\!_R}\!\!> \beta_{_{\rm b}}$ corresponding to 
when electrons see most incoming photons with $\theta_{_R}\approx0$, 
and deceleration ($f_z<0$) for  $\cos\tilde{\theta}_{_R}<\beta_{_{\rm b}}$ 
corresponding to when electrons see most photons with 
$\theta_R\approx\pi$ in the jet frame. Therefore, the key role in 
radiative acceleration or deceleration is played by anisotropic 
scattering, which can be understood as follows. Assume a cell moving 
along the jet containing relativistic plasma with an isotropic 
distribution. The particles in the jet frame see anisotropic 
photon fields, i.e. the radiation mainly comes from behind or in
front depending on $\Gamma$ of the bulk motion. When 
$\cos\tilde{\theta}_{_R}<\beta_{_{\rm b}}$, in the jet frame, 
although each electron would scatter photons into its direction of
motion, on average there is more momemtum beamed along the jet 
direction. Thus, the cell is subject to a force in the opposite 
direction.  The critical $\Gamma$, above which the radiative 
deceleration ($f_z<0$) occurs, can be derived from (2.6) by setting 
$f_z=0$ (e.g. Sikora et al. 1996). Since the critical $\Gamma$ is 
generally small (e.g. Phiney 1987; Sikora et al. 1996), our 
discussion concentrates on deceleration of jets with a large initial 
$\Gamma$. 

In the case of deceleration, since the flux (2.2) is peaked at 
$R=R_0=(9/7)R_{_{\rm min}}$, an approximation for $f_z$ at $z\gg R_0$
is given by
$$\eqalign{f_z\approx&-{\textstyle{1\over3}}\mu_{_p}\eta\Gamma^2
\,{R_gc\over z^2}\,\Bigl({R_0\over z}\Bigr)^4\langle\gamma^2\rangle
n_{e0}\cr
=&-
{\textstyle{2\over3}}\mu_{_p}\eta\Gamma^2
\,{R_gc\over z^2}\,\Bigl({R_0\over z}\Bigr)^4
\gamma^2_{_{\rm min}}
n_{e0}\,\ln(\gamma_{_{\rm max}}/\gamma_{_{\rm min}}),
\cr},\eqno(2.8)$$ 
where $\mu_{_p}=m_p/m_e$ is the proton to electron mass ratio, 
$R_0=7R_{_{\rm min}}/6$ is the radius at which the flux is peaked, 
$\eta=L_d/L_{_{\rm Edd}}$ is the radiation efficiency of disc emission,
$L_{_{\rm d}}=2\pi m_ec^2\int dR\,RF(R)\tilde{\varepsilon}_{_R}$ is 
the luminosity of the source, and the equality is for $p=3$. 
The radiation force increases rapidly with $\Gamma^2$.

Although (2.6) is derived using the beaming approximation as described
by $\delta(\Omega_s-\Omega_e)$ in Eq. (A1), apart from $\beta\approx1$ in 
the average $\langle\gamma^2\beta\rangle$, it is in good agreement with 
the result derived by Sikora et al. (1996) using covariant formalism. One 
may also reproduce the result given by O'Dell (1981) where the photons are 
from a point source, for which we may assume $\tilde{\theta}_{_R}=0$ and 
that $2\pi\int dR\,RF(R)\tilde{\varepsilon}_{_R} \to L_{_{\rm d}}/m_ec^2$ 
is the luminosity of the source. In this special case, the particles are 
accelerated away from the source.

In calculating (2.6) and (2.8), we assume $\varepsilon'<1$. In the electron
rest frame, the photon energy is $\varepsilon'=\varepsilon'_{_R}\equiv
\gamma (1-\beta\cos\Theta_{_R})\Gamma(1-\beta_{_b}\cos\tilde{\theta}_{_R})
\tilde{\varepsilon}_R$, where $\Theta_R$ is the angle between the incoming
photon direction and electron motion (in $K_j$). Then, the condition
for {\it all} particles to scatter in the Thomson regime is
$$\gamma\varepsilon_{_R}<1/2. \eqno(2.9)$$
with $\varepsilon_{_R}=
\Gamma(1-\beta_b\cos\tilde{\theta}_R)\tilde{\varepsilon}_{_R}$.
This condition requires that photons of the highest energy, seen by 
particles with $\Theta_{_R}=\pi$, satisfy $\varepsilon'_{_R}<1$, which 
may not be the case if $\Gamma$, or the average Lorentz factor 
($\langle\gamma\rangle$) of the plasma in $K_j$, are large.

\subsection{The Klein-Nishina regime}

For an ultrarelativistic jet with a large bulk Lorentz factor 
or when particles are highly relativistic in the jet comoving frame,
Compton scattering should be treated in the Klein-Nishina scattering 
regime.  The minimum $\Gamma$ such that {\it all} particles of energy 
$\gamma$ are in the KN regime can be derived as follows. In the jet 
frame, since the soft photon direction is $\theta_R\approx\pi$, 
electrons with $\theta_e=\pi$ see the lowest energy photons. Since
$\cos\Theta_R=\cos\theta_R\cos\theta_e+\sin\theta_R\sin\theta_e\cos(\phi-
\phi_e)\approx-\cos\theta_{_R}\cos\theta_e$ (where $\phi$ and $\phi_e$ 
are the azimuthal angles of the photon and electron), and from 
$\varepsilon'_{_R}\geq1$, we can derive the condition for all 
particles to scatter in the KN regime, 
$${\Gamma\over\gamma}>{2\over\tilde{\varepsilon}_R(1-\beta_{_b}
\cos\tilde{\theta}_R)}, \eqno(2.10)$$
or
$${\Gamma\over\gamma}<{\textstyle{1\over2}}\tilde{\varepsilon}_{_R}
(1+\cos\tilde{\theta}_{_R}), \eqno(2.11)$$
where we assume that the plasma is relativistic in the jet frame.  
For $\tilde{\varepsilon}_R=6\times10^{-5}$, we have $\Gamma/\gamma
>5\times10^4/(1-\cos\tilde{\theta}_R)$ or $\Gamma/\gamma<
3\times10^{-5}(1+\cos\tilde{\theta}_R)$. 
Inequalities (2.10) and (2.11) are the KN condition
for particles of $\gamma$ with any $\theta_e$, which requires 
an extreme ratio (very large or very small) $\Gamma/\gamma$. However,
in the following discussion we show that the condition for
the KN scattering to be the dominant process in radiative
drag is less strict than (2.10), (2.11), and is generally 
satisfied for moderate $\Gamma$.

\beginfigure{3}
\psfig{file=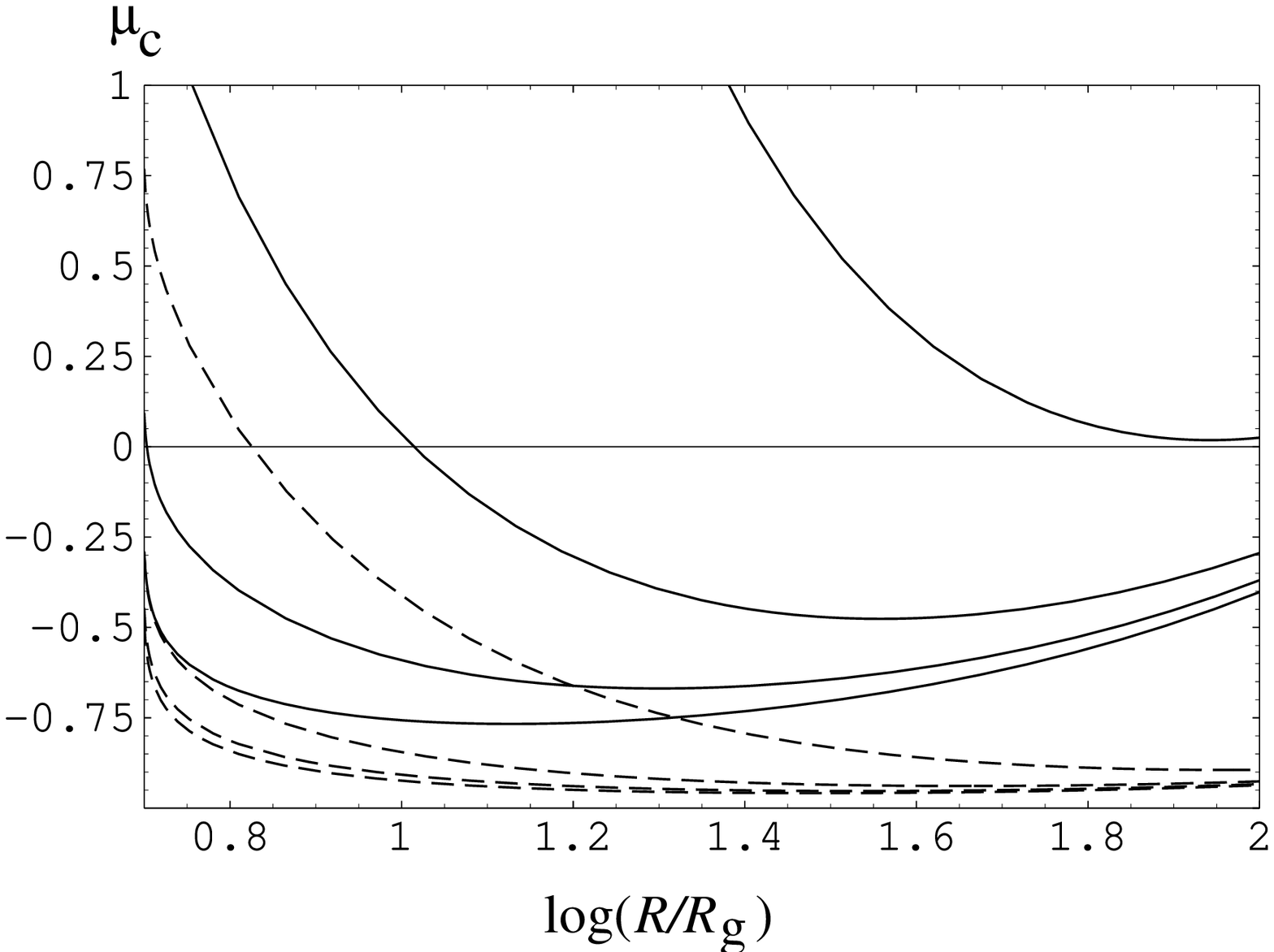,height=7cm,width=8cm}
\caption{{\bf Figure 3.}
Plots $\mu_c$ as a function of $R/R_g$ for different $z$ and
$\gamma$. The solid and dashed curves correspond respectively to
$\gamma=10^2$ and $10^3$. In each case, the curves from bottom to top
correspond to $z/R_g=5$, 10, 30, 50. We assume the bulk Lorentz factor 
$\Gamma=10^4$.
}
\endfigure

\beginfigure{4}
\psfig{file=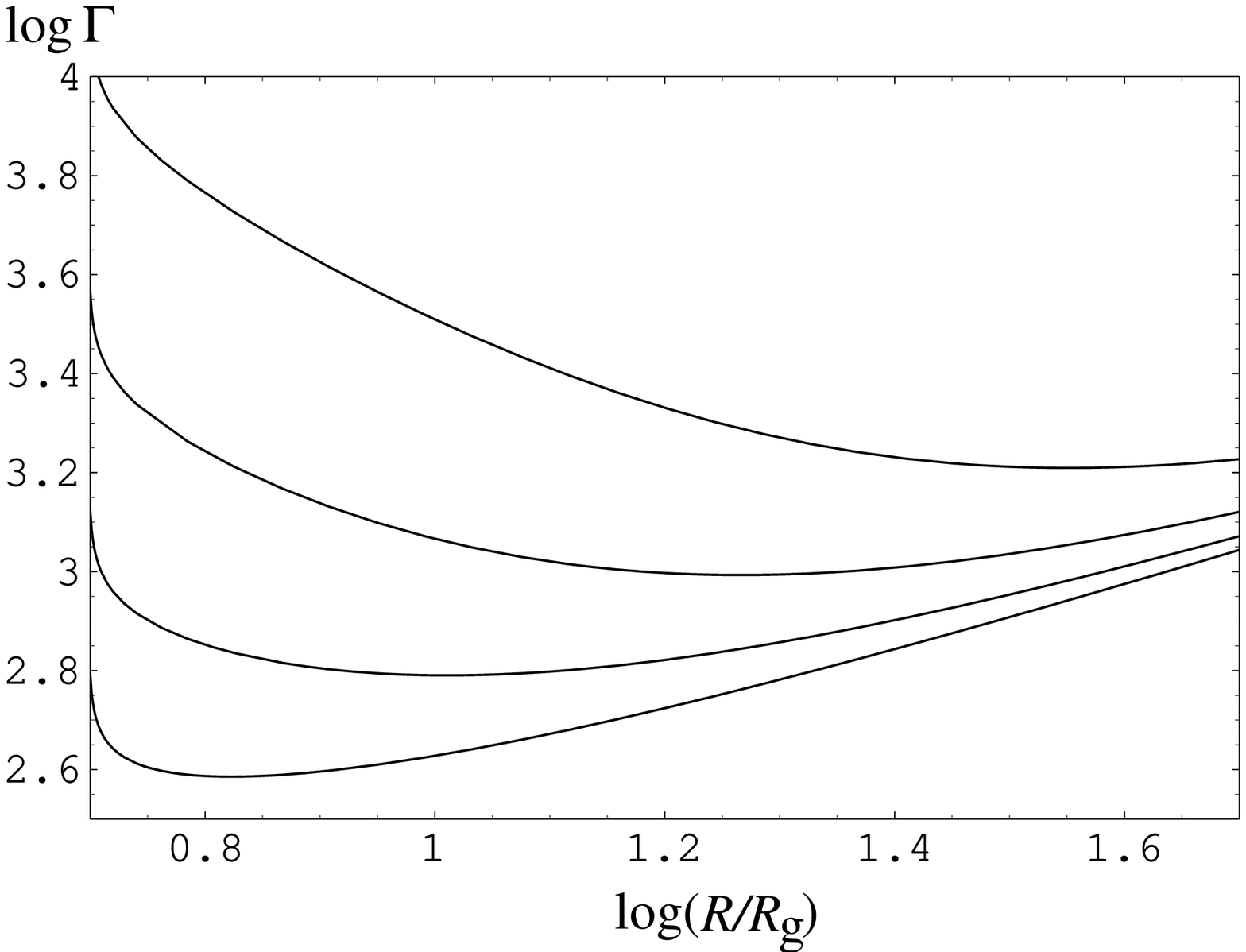,height=7cm,width=8cm}
\caption{{\bf Figure 4.} The minimum $\Gamma$ for KN scattering to be 
dominant in deceleration are plotted as functions of $R/R_g$. The curves
from top to bottom correspond to $z/R_g=20$, 10, 5, 2. We assume 
$\gamma_{_{\rm min}}=80$, $R_{\rm min}=5R_g$, and use the temperature
function (2.3) with $\dot{M}=0.1L_{\rm Edd}/c^2$.
}
\endfigure

In general, given a particle distribution, the Thomson and KN
approximations apply respectively to particles with $\cos\theta_e<\mu_c$
and $\cos\theta_e\geq\mu_c$, where $\mu_c$ satisfies
$\Gamma\gamma(1-\beta\mu_c\cos\theta_R)(1-\beta_b\cos\tilde{\theta}_R)
\tilde{\varepsilon}_R=1$, that is,
$$\mu_c=\cases{\displaystyle{{1\over\beta}
\,{1-\Gamma\gamma(1-\beta_b\cos\tilde{\theta}_R)\tilde{\varepsilon}_R\over
\Gamma\gamma(\beta_b-\cos\tilde{\theta}_R)\tilde{\varepsilon}_R}},
&for $\gamma\varepsilon_{_R}\geq1/2$;\cr
\displaystyle{1},&for $\gamma\varepsilon_{_R}<1/2$.\cr}\eqno(2.12)$$
Thus, the conditions (2.10), (2.11) correspond to an extreme case; if one 
of these two conditions is satisfied, then $\mu_c=-1$, and in this 
case the KN approximation applies to {\it all} particles. Figure 3 shows 
plots of $\mu_c$ as functions of $R/R_g$. For given $z$ and $\gamma$, 
the KN scattering regime is located above the curve (i.e. for electrons 
or positrons with $\cos\theta_e\geq\mu_c$). 

To calculate $f_z$, particles of energy $\gamma$ in the cell are divided 
into two parts according to (2.12). Let $f_{z,{\rm T}}$ and $f_{z,{\rm KN}}$
be the force due to scattering by particles with $\cos\theta_e\leq
\mu_c$ (the Thomson regime) and $\cos\theta_e>\mu_c$ (the KN regime), 
respectively. Then, the radiation force can be expressed as 
$$f_z=f_{z,{\rm T}}+f_{z,{\rm KN}}.\eqno(2.13)$$
where $f_{z,{\rm T}}$ and $f_{z,{\rm KN}}$ can be calculated approximately
using the cross sections (A1), (A2) as given in the Appendix (cf. Eq. A4 
and A5).  Assuming that incident photons in the jet frame are 
approximately beamed, $\cos\theta_{_R}\approx-1$, which is valid for 
$\Gamma\gg1$,  we have
$$\eqalign{f_{z,{\rm T}}&=-{\textstyle{1\over2}}\sigma_{_{\rm T}}
C_0n_{e0}\!\!
\int^{R_{_{\rm max}}}_{R_{_{\rm min}}}\!\!\!dR\,
{RF\tilde{\varepsilon}_{_R}\Gamma^2\over
2\pi(R^2+z^2)}(1-\beta_b\cos\tilde{\theta}_{_R})^2\cr
\times&\!\!\int^{\gamma_{_{\rm max}}}_{\gamma_{_{\rm min}}}\!\!\!d\gamma\,
\gamma^{-p}\Bigl\{\gamma^2\Bigl[
{\textstyle{1\over2}}(\mu^2_c\!-\!1)+{\textstyle{2\over3}}(\mu^3_c\!+\!1)
+{\textstyle{1\over4}}(\mu^4_c\!-\!1)\Bigr]\cr
+&(\mu_c+1)+{\textstyle{1\over2}}(\mu^2_c-1)\Bigr\},
\qquad\qquad\qquad\qquad\qquad\qquad\llap{(2.14)}\cr}
$$
and
$$\eqalign{f_{z,{\rm KN}}&\approx-{\textstyle{3\over8}}\sigma_{_{\rm T}}
C_0n_{e0}\!\int^{R_{_{\rm max}}}_{R_{_{\rm min}}}\!\!\!dR\,
{RF\over2\pi(R^2+z^2)}\,{1\over4\tilde{\varepsilon}_{_R}}\cr
\times&\!\!\int^{\gamma_{_{\rm max}}}_{\gamma_{_{\rm min}}}
\!d\gamma\,\gamma^{-p}\Bigl\{
\bigl(1-\mu^2_c\bigr)\ln2+(1-\mu_c)
\cr
+{2\varepsilon_{_R}\over\gamma}&
\Bigl[(1-\mu_c)\ln(2/\sqrt{e})
+2\ln(2\varepsilon_{_R}\gamma)\Bigr]\!\Bigr\}
H(\gamma\varepsilon_{_R}\!-\!{\textstyle{1\over2}}),\cr}
\eqno(2.15)$$
where $H(\gamma\varepsilon_{_R}-1/2)=1$ for $\gamma\varepsilon_{_R}\geq1/2$
and $H(\gamma\varepsilon_{_R}-1/2)=0$ for $\gamma\varepsilon_{_R}<1/2$.
In the Thomson approximation, $\mu_c=1$, 
we have $f_z=f_{z,{\rm T}}$, which reduces to (2.6) with 
$\cos\theta_{_R}\approx-1$. For $\mu_c<1$, $f_{z,{\rm T}}$ can be
positive even though we assume $\cos\theta_{_R}=-1$. This is because
$f_{z,{\rm T}}$ includes particles with $-1\leq\cos\theta_e<\mu_c$,
and if most of them have $\theta_e>\pi/2$ we would have $f_{z,{\rm T}}>0$. 
In contrast, $f_{z,{\rm KN}}$ is always negative since it includes only 
those particles with $\mu_c<\cos\theta_e\leq1$, most of which move
forward along the jet and contribute to drag. From (2.14) and (2.15), 
we see that when $\mu_c<1$ scattering in the KN regime becomes 
important while the force due to scattering in the Thomson regime 
starts to decrease. From (2.12), this condition becomes
$$\Gamma\geq
{1\over2\gamma(1-\beta_{_b}\cos\tilde{\theta}_{_R})
\tilde{\varepsilon}_{_R}}.
\eqno(2.16)$$
Figure 4 shows the condition (2.16) as a function of $R$ with
$\gamma$ being replaced by $\langle\gamma\rangle$. We assume 
$\gamma =80$, $p=3$. Thus, the average Lorentz factor in the jet frame is
$\langle\gamma\rangle=160$. This gives, for example, $\Gamma>400$ 
for $z=2R_g$, and $\Gamma>700$ for $z=5R_g$. These lower limits can be
further reduced as scattering in the KN regime extends to higher
$\gamma$.

\beginfigure{5}
\psfig{file=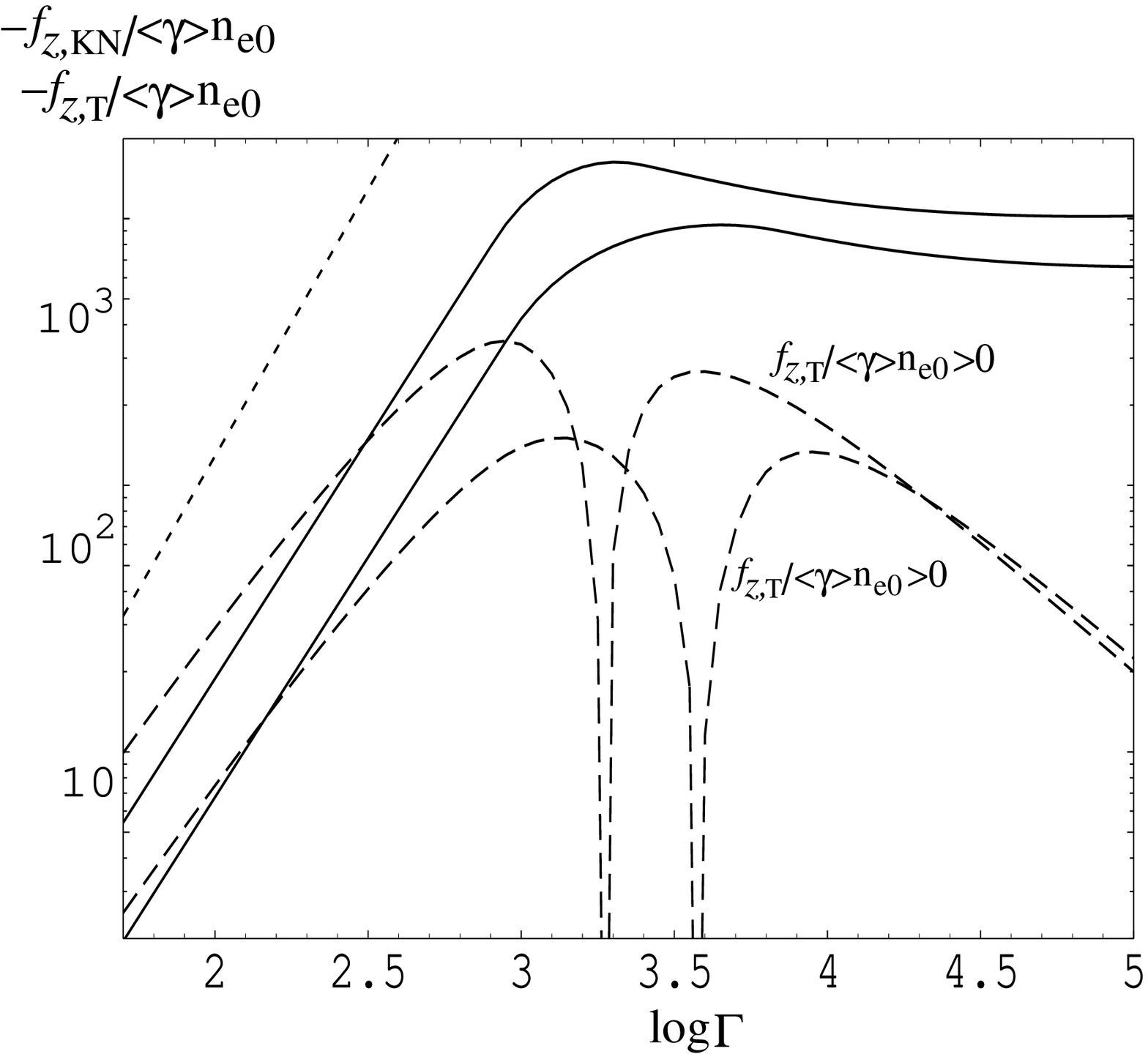,height=6cm,width=9cm}
\medskip
\caption{{\bf Figure 5.} Variation of $f_{z,{\rm T}}/\langle\gamma\rangle
n_{e0}$ (dashed), $f_{z,{\rm KN}}/\langle\gamma\rangle n_{e0}$
(solid) with $\Gamma$ for $z=5R_g$ (upper) and $z=10R_g$ (lower).
The dotted curve (for $z=5R_g$) is calculated by ignoring scattering
in the KN regime.
}
\endfigure

Integration over $\gamma$ in Eq. (2.14) and (2.15) can be done
analytically using (2.7), and is given by Eq. (A8), (A9) in the Appendix. 
For $2\varepsilon_{_R}<1/\gamma_{_{\rm min}}$, (2.15) has the following
approximation
$$\eqalign{&f_{z,{\rm KN}}\approx\cr
&-\alpha\sigma_{_T}n_{e0}
\!\!\int^{R_{_{\rm max}}}_{R_{_{\rm min}}}\!\!\!dR\,
{RF(R)\over2\pi(R^2+z^2)}\,{3(\varepsilon_{_R}\gamma_{_{\rm min}})^{p-1}
\over16\tilde{\varepsilon}_R},\cr}
\eqno(2.17)$$  
with $\alpha=2/(p-1)+\ln(64e)/p-\ln2/(p+1)$, which ranges from
4.4 ($p=2$) to 2.6 ($p=3$). For large $\Gamma$ satisfying 
(2.16), $f_{z,{\rm KN}}$ overtakes $f_{z,{\rm T}}$ and become 
important. For even larger $\Gamma$, we find 
$$\eqalign{f_z\approx f_{z,{\rm KN}}=&
-{\textstyle{3\over8}}\sigma_{_{\rm T}}n_{e0}\!
\int^{R_{_{\rm max}}}_{R_1}\!dR\,
{RF\over2\pi(R^2+z^2)}\cr
\times&{1\over2\tilde{\varepsilon}_{_R}}
\Bigl[1+{2\varepsilon_{_R}(p-1)\over\gamma_{_{\rm min}}p}\,
\ln(2\varepsilon_{_R}\gamma_{_{\rm min}})\Bigr],\cr}
\eqno(2.18)$$
where $R_1$ is the radius of the ring which emits photons with energy 
$\varepsilon_{_R}=1/(2\gamma_{_{\rm min}})$. If 
$\varepsilon_{_R}>1/(2\gamma_{_{\rm min}})$ for $R_{_{\rm min}}\leq R\leq
R_{_{\rm max}}$, $R_1$ is replaced by $R_{_{\rm min}}$. Figure 5 shows 
$f_{z,{\rm T}}/\langle\gamma\rangle n_{e0}$ (Eq. 14), 
$f_{z,{\rm KN}}/\langle\gamma\rangle n_{e0}$ (Eq. 15) 
as functions of $\Gamma$ for $z=5R_g$, $10R_g$. The disc
parameters are the same as in Figure 2. We assume $p=2.8$,
$R_{_{\rm min}}=5R_g$, $R_{_{\rm max}}=50R_g$. The value of $f_z$ is
significantly lower than that calculated using the Thomson approximation
(dotted curve), indicating the importance of the KN scattering. 
The KN  scattering starts to overtake the Thomson scattering when
$\Gamma>300$ for $z=5R_g$ and $\Gamma>200$ for $z=10R_g$, and it is
described by Eq. (2.17).  Scattering in the KN regime becomes dominant 
in decelerating the jet at $\Gamma>10^3$  and the approximation (2.18)
applies.  The force component due to scattering in the Thomson regime 
becomes positive for large $\Gamma$ since most particles that scatter in the 
Thomson regime move toward the black hole and hence contribute to
acceleration of the cell. However, the net force is still negative
because the component due to forward moving particles that scatter in 
the KN regime dominates. For extremely large $\Gamma$, we may have
$2\varepsilon_{_R}>p\gamma_{_{\rm min}}/(p-1)\ln(2\varepsilon_{_R}
\gamma_{_{\rm min}})$, and then, the second term on the right hand side
of (2.18) becomes important, and $f_z\approx f_{z,{\rm KN}}$ increases 
logarithmically with $\Gamma$.

\section{Radiative deceleration}

A relativistic jet can form in the vicinity of a black hole through
either (1) a hydrodynamic process, in which radiation from an accretion
disc plays an important role in producing and collimating the
jet, or (2) a MHD process, in which magnetic fields play major role
in the jet flow. In either case, regardless of the details of jet 
production, which are poorly understood, the jet must pass through 
intense radiation fields from disc emission and be subject to radiative
deceleration. In the following discussion, we assume that the jet has 
initially a large $\Gamma$ and then undergoes deceleration due to the 
radiation force discussed in earlier sections, and we examine the constraint 
on the bulk flow when scattering in the KN regime is included.

\subsection{Constraint on $\Gamma$}

The bulk Lorentz factor $\Gamma$ as a function of $z$ can be calculated
by integrating (2.1) along the jet direction. For deceleration from 
the KN to Thomson regime, we use (2.13) together with (2.14) and (2.15) 
for which integration over $R$ can be done numerically. We first consider 
an electron positron jet with an energy density (in $m_ec$) given by
$E=n_{e0}\langle\gamma\rangle$. (A relativistic jet containing protons
will be considered in Sec.3.2.) From the distribution (2.7), we have
$$E\approx n_{e0}\gamma_{_{\rm min}} {p-1\over p-2}\,
\Biggl[1-\Biggl({\gamma_{_{\rm min}}\over\gamma_{_{\rm max}}}\Biggr)^{p-2}
\Biggr],\eqno(3.1)$$
for $p>2$ and $\gamma_{\rm max}\gg\gamma_{\rm min}$.
Figure 6 shows $d\ln\Gamma/dz$ as a function of $z$ and $\gamma$. Radiative 
drag is severe, i.e. $d\ln\Gamma/dz \ll-1$, for small $z$ and moderate $\Gamma$, 
but decreases as $z$ increases. Plots of $d\ln\Gamma/dz$ as a function of
$z$ for an electron-positron jet are given in the solid curves in 
Figure 7. For a very large bulk Lorentz factor, the deceleration efficiency
decreases significantly because scattering is in the KN regime. For
example, we have a much smaller $d\ln\Gamma/dz$ for $\Gamma=10^5$, as
shown in Figure 7. 
\beginfigure{6}
\psfig{file=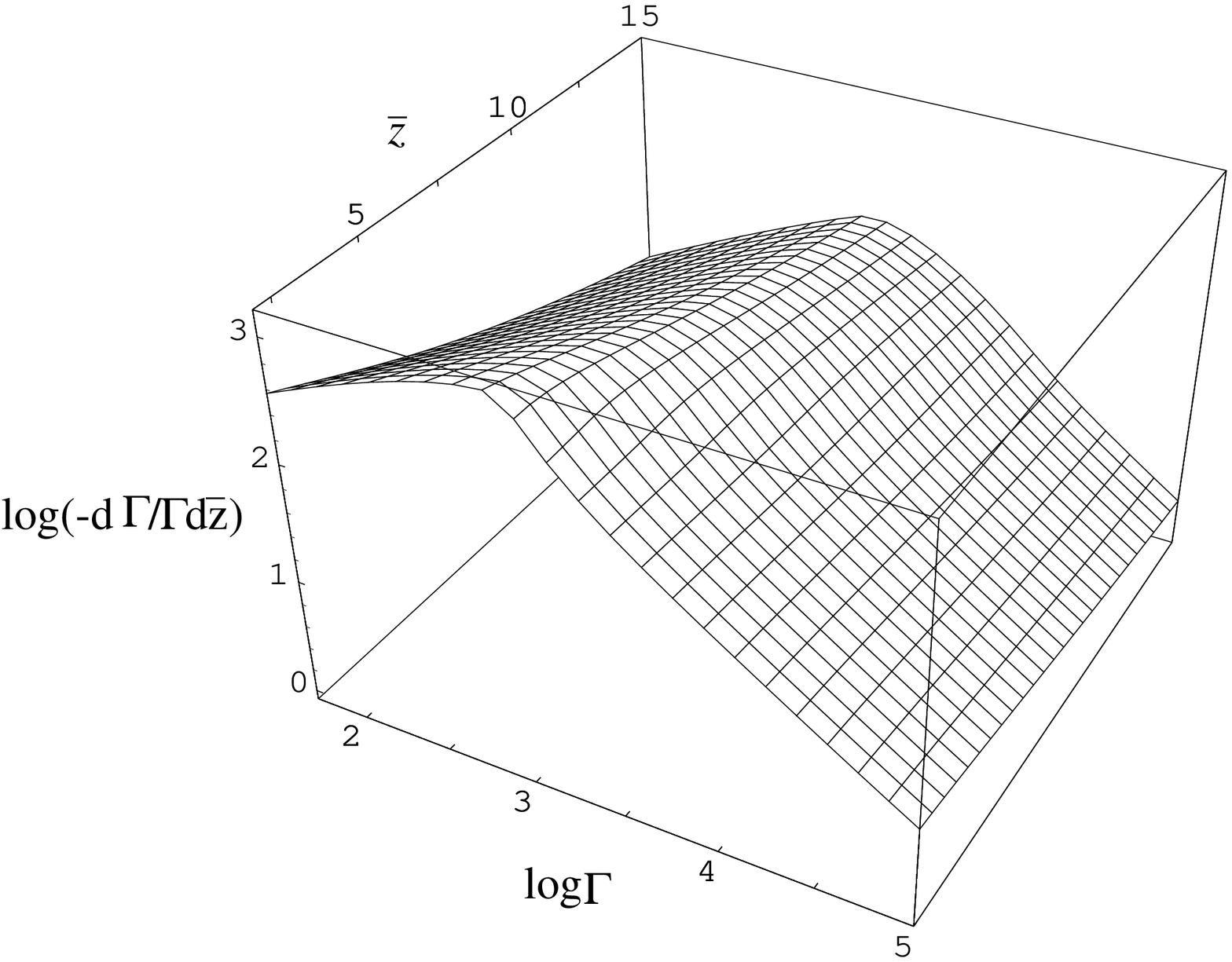,height=6cm,width=7cm}
\medskip
\caption{{\bf Figure 6.} A plot of $-d{\rm ln}\Gamma/d\bar{z}$ vs. $\bar{z}$
(with $\bar{z}=z/R_g$) and $\Gamma$. The disc parameters are the same as 
in Figure 5.
}
\endfigure

\beginfigure{7}
\psfig{file=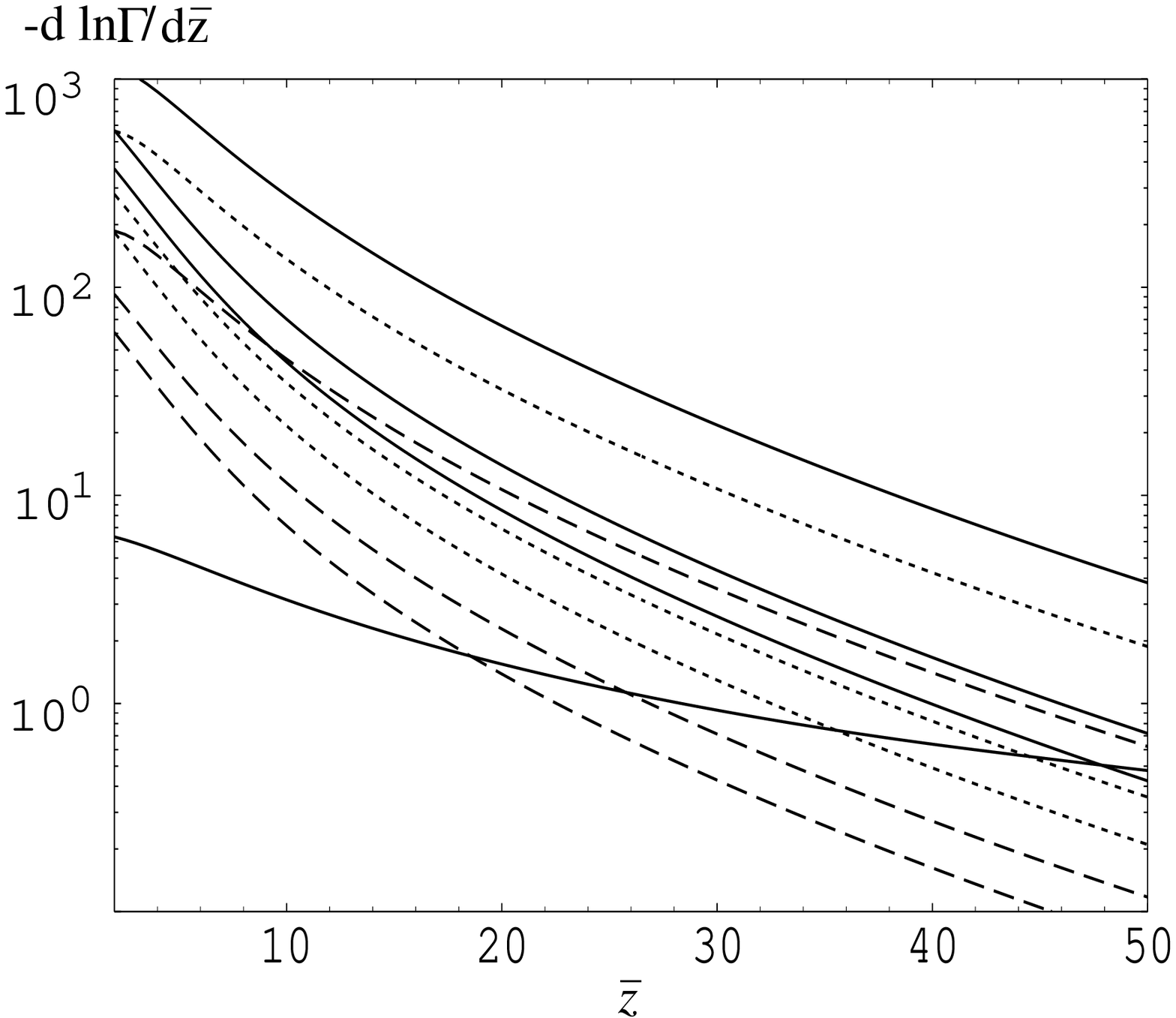,height=7cm,width=9cm}
\medskip
\caption{{\bf Figure 7.} Plots of $-d{\rm ln}\Gamma/d\bar{z}$ vs. $\bar{z}$
($\bar{z}=z/R_g$). The solid curves, from bottom to top corresponding to
$\Gamma=10^5$, $50$, $10^2$, $10^3$, are for a pure electron-positron 
jet (i.e. $n_p/n_{e0}=0$, cf. Eq. 3.1). The dotted and dashed curves are 
for the jet containing cold protons with $n_p/n_{e0}=0.1$ and $0.5$, 
respectively (cf. Sec. 3.2). In each of these two cases, the curves from 
bottom to top correspond to $\Gamma=50$, $10^2$, $10^3$. The disc parameters 
are the same as in Figure 5.
}
\endfigure

In the Thomson limit, we can solve (2.1) with (2.8) to find
$$\Gamma(z)\approx
{\Gamma_0\over1+\eta\Gamma_0\xi_{_T}},\eqno(3.2)$$
where $\Gamma_0$ is the initial value at $z=z_0$, and
$$\xi_{_T}(z_0,z)={\textstyle{1\over15}}
\mu_{_p}{R_g\over z_0}\Bigl({R_0\over z_0}\Bigr)^5
\Bigl(1-{z^5_0\over z^5}\Bigr)
{\langle\gamma^2\rangle\over \langle\gamma\rangle}.
\eqno(3.3)$$
For $p=3$ and the distribution given by (2.7), $\xi_{_T}$ reduces to
$$\xi_{_T}\approx10^2{R_g\over z_0}
\Bigl({R_0\over z_0}\Bigr)^5
\Bigl(1-{z^5_0\over z^5}\Bigr)
\gamma_{_{\rm min}}\ln(\gamma_{_{\rm max}}/\gamma_{_{\rm min}}).\eqno(3.4)$$
Assuming that the initial value of $\Gamma$ is large and
$$\eta\Gamma_0\,\xi_{_T}\gg1,$$
one has the asymptotic value given by
$$\Gamma (\infty)\approx
{1\over\eta\,\xi_{_T}(z_0,\infty)}.\eqno(3.5)$$
For, $p=3$, $\gamma_{_{\rm max}} /\gamma_{_{\rm min}}=5\times10^3$,
$\gamma_{_{\rm min}}=50$, $R_{_{\rm min}}=2R_g$, and $z_0=10R_g$,
even with $\Gamma_0\gg1$, we have $\Gamma(\infty)\approx 1$ for
$\eta=0.1$, and $\Gamma(\infty)\approx 10$ for $\eta=0.01$. Thus, in
the Thomson limit,  radiation drag is indeed a severe constraint on
$\Gamma$ as already shown in earlier work, e.g. by Melia \& K\"ongl
(1989), Sikora et al. (1996).

Deceleration described by (3.2) applies only for small $\Gamma$ or 
$\gamma_{_{\rm max}}$. When $\Gamma$ satisfies (2.16), 
deceleration is mainly due to particles scattering in the KN regime. 
From (2.17), and using (2.1), the bulk Lorentz factor is calculated to be
$$\Gamma(z)\approx
{\Gamma_0\over[
1+\eta\Gamma^{(p-2)}_0\xi_{_{\rm KN}}]^{1/(p-2)}},
\eqno(3.6)$$
with 
$$\eqalign{&\xi_{_{\rm KN}}=\cr
&
{3\alpha(p-2)^2\mu_{_p}\tilde{\varepsilon}^{p-3}_{_R}\over 
16\pi 2^{p-1}(2p-1)}
{R_g\over R_0}\Bigl(
{R_0\over z_0}\Bigr)^{2p-1}\Bigl[1-\Bigl({z_0\over z}\Bigr)^{2p-1}
\Bigr]\gamma^{p-2}_{_{\rm min}},\cr}\eqno(3.7)$$
where $p>2$, $\alpha=4.4-2.6$, $\tilde{\varepsilon}_{_R}=5\times10^{-5}$.
Deceleration is significantly slower than that calculated from
the force in the Thomson approximation since 
$\xi_{_{\rm KN}}$ is smaller than $\xi_{_{\rm T}}$ by a factor of 
$5\gamma_{_{\rm min}}\ln(\gamma_{_{\rm max}}/\gamma_{_{\rm min}})$
(for $p=3$). Eq. (3.6) applies only within a narrow range of $\Gamma$
(i.e. $1/\langle\gamma\rangle\leq\varepsilon_{_R}\leq1/\gamma_{_{\rm min}}$)
as shown in Figure 5. 

For even larger $\Gamma_0$, the drag is mainly due to scattering in
the KN regime (2.18). Retaining only the first term on the right hand 
side of (2.18), we have 
$$\Gamma(z)=\Gamma_0-{3\eta\mu_{_p}\over16\pi\langle\gamma\rangle
\tilde{\varepsilon}^2_{_R}}\,{R_g\over z_0}\,\Bigl(1-{z_0\over z}\Bigr),
\eqno(3.8)$$
for $z_0\leq z\leq z_1$, where $z_1$ is the distance at which $\Gamma(z_1)$
is so small that (3.8) does not apply. Figure 8 shows the rapid drop of
$\Gamma(z)$ with the distance $z$, where we assume that
the jet starts with $\Gamma_0=10^5$ at $z_0=5R_g$ and $10R_g$,
respectively. In each case, the first curve from the left corresponds
to the deceleration of an $e^\pm$ jet.  The solid curves correspond to 
deceleration
in the Thomson regime, and are calculated using (2.6). The bulk Lorentz
factor decreases from $\Gamma=10^5$ to $\Gamma=10^3$ within a distance
$\Delta z\ll z_0$. Radiation drag is effective except for 
$\Gamma_0\geq (3\eta\mu_{_p}/
16\pi\langle\gamma\rangle\tilde{\varepsilon}^2_{_R})(R_g/z_0)
\approx5\times10^6(\eta/0.1)(R_g/z_0)$ (for $\langle\gamma\rangle=100$). 
However, for such a large $\Gamma_0$, the second term becomes 
important, and $\Gamma$ decreases exponentially according to
$$\eqalign{&\ln(\Gamma/\Gamma_0)\approx\cr
&-{3\eta\mu_{_p}(p-1)\over4\pi p}\int^z_{z_0}\!dz'\,
{R_g\over {z'}^2}\,{1-\cos\tilde{\theta}_{_R}\over
\tilde{\varepsilon}_{_R}\langle\gamma\rangle
\gamma_{_{\rm min}}}\ln(2\varepsilon_{_R}\gamma_{_{\rm min}}).
\cr}\eqno(3.9)$$ 
For $p=3$, we have
$$\ln(\Gamma/\Gamma_0)\approx
-5.3\Bigl({\eta\over0.1}\Bigr){R_gR^2_0\over z^3_0}
\Bigl(1-{z^3_0\over z^3}\Bigr)
{\ln(2\varepsilon_{_R}\gamma_{_{\rm min}})\over
\tilde{\varepsilon}_{_R}\gamma^2_{_{\rm min}}},\eqno(3.10)$$
with $\varepsilon_{_R}=\Gamma_0\tilde{\varepsilon}_{_R}$.
The drag strongly depends on $\gamma_{_{\rm min}}$, and can be
reduced for large $\gamma_{_{\rm min}}$.

\beginfigure{8}
\psfig{file=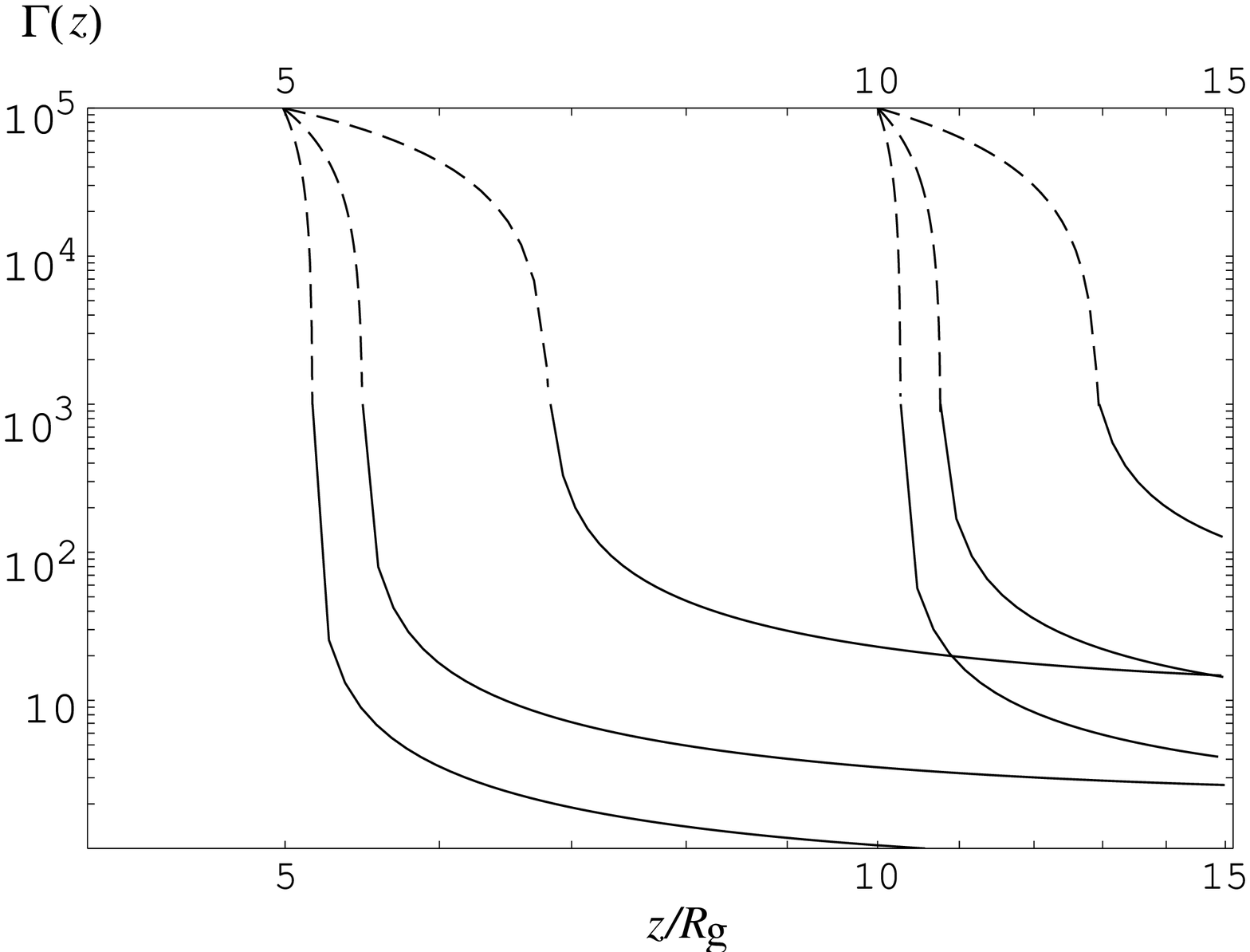,height=7cm,width=8cm}
\medskip
\caption{{\bf Figure 8.} Deceleration of jets. Deceleration in
the KN regime and Thomson regime are indicated respectively  by 
dashed and solid part of the curves. The jet is assumed to start 
with $\Gamma=10^5$ at $z=5R_g$ and $z=10R_g$. In each case, the
curves from left to right correspond to $n_p/n_{e0}=0$ (a pure
electron-positron jet), $n_p/n_{e0}=0.1$, and $n_p/n_{e0}=0.5$.
The efficiency of disc emission, $\eta=0.05$, is used. 
Protons are assumed to be cold.
}
\endfigure

\subsection{A jet containing protons}

If a jet is heavily loaded with protons, the effect of radiation drag
is reduced significantly, since protons scatter photons with 
much a smaller cross section compared to electrons, and the presence 
of protons effectively increases the inertia. When the jet contains 
cold protons, the energy density is $E=2n_{e0}\gamma_{_{\rm min}}+n_p\mu_{_p}$
where $n_p$ is the number density of protons and the electron spectral
index is assumed to be $p=3$ (cf. Eq. 2.7). Suppose that $n_p$ is so 
large that the condition 
$$\delta\equiv{\mu_{_p}n_p\over2\gamma_{_{\rm min}} n_{e0}}>1\eqno(3.11)$$ 
is satisfied. The value $\delta$ in Eq. 3.11, corresponding to the ratio 
of the cold proton energy density to the electron energy density, can 
increase significantly if protons are relativistic in the jet frame.  
Then, $f_z/E$ is reduced by a factor $\delta$.  The asymptotic 
$\Gamma(\infty)$ given by (3.5) would increase by a factor $\delta$. For 
example, for $\gamma_{_{\rm min}}=50$, one must include the inertia of 
protons if $n_p/n_{e0}>0.05$, and the drag force would be significantly 
reduced.

The effect of protons on the deceleration efficiency is shown in
Figure 7 by dotted curves for $n_p/n_{e0}=0.1$ and dashed curves for
$n_p/n_{e0}=0.5$. We have smaller $d\ln\Gamma/dz$ than that for the
$e^\pm$ jet, indicating a decrease in the deceleration efficiency. 
A relatively weaker radiation drag on jet containing protons
can also be seen in Figure 8. The jets with $n_p/n_{e0}=0.1$ and 0.5
have much higher terminal Lorentz factors than the pure $e^\pm$ jet.

\subsection{Energy loss due to inverse Compton scattering}

In the calculation of the radiation force, we assume that the plasma 
in the jet is highly relativistic. In practice, the particle distribution
is determined by the various energy loss processes and the injection rate
of high energy particles. The energy loss rate due to
inverse Compton scattering can be derived 
from $-\dot{\gamma}=-{\bf \beta}\cdot d{\bf p}/dt$ with
$d{\bf p}/dt$ given by (2.5).  In the Thomson limit, the loss rate
is obtained in the jet frame ($K_j$) as
$$\eqalign{-\langle\dot{\gamma}\rangle_a\!=\!&\!\!
\int^{R_{\rm max}}_{R_{\rm min}}\!\!\!dR\,{RF(R)\over R^2+z^2}\,
\Gamma^2\,(1\!-\!\beta_b\cos\tilde{\theta}_R)^2{4\sigma_{_T}\over3}\gamma^2
\tilde{\varepsilon}_{_R}\cr
&\approx {\textstyle{2\over3}}\mu_{_p}\eta{
cR_g\over z^2}\biggl(
{R_0\over z}\biggr)^4\Gamma^2\gamma^2,\cr}\eqno(3.12)$$
where $\langle...\rangle_a$ represents an average over $\Omega_e$,
and the approximation is valid only for $\Gamma\gg z/R_0$.
In the large angle approximation, we have 
$-\langle\dot{\gamma}\rangle_a\propto\gamma^2\Gamma^2$.  In the 
small angle approximation $\tilde{\theta}_{_R}\ll \Gamma^{-1}$, i.e.
the photon field is from a point source, we have $-\langle\dot{\gamma}
\rangle_a\propto\gamma^2/\Gamma^2$. From (3.12), the time scale in $K$
for energy loss for $\Gamma\gamma<1/\tilde{\varepsilon}_{_R}$ 
and $z>R_0$ is estimated to be
$$\eqalign{t_{_{\rm IC}}&
\approx{\textstyle{3\over2}}{z^2\over cR_g}
\biggl({z\over R_0}\biggr)^4{1\over \Gamma\gamma \eta}\cr
=& 0.017\,{\rm s}\,\Bigl({0.1\over \eta}\Bigr)
\biggl({\tilde{\varepsilon}_{_R}
\over 6\times10^{-5}}\biggr)
\biggl({1/\tilde{\varepsilon}_{_R}
\over\Gamma\gamma}\biggr) \biggl({z\over 10R_g}\biggr)^6,\cr}
\eqno(3.13)$$
where $\tilde{\varepsilon}_{_R}=6\times10^{-5}$ corresponds to 30 eV
photons. For $\Gamma\gamma\geq 1/\tilde{\varepsilon}_{_R}$, the 
time scale should be calculated in the KN limit.

In the KN scattering regime, their energy loss rate is much slower, i.e.
$$\eqalign{-\langle\dot{\gamma}\rangle_a&\approx\int
^{R_{\rm max}}_{R_{\rm min}}\!dR\,{RF(R)\over R^2+z^2}\,
{3\sigma_{_T}\over8\tilde{\varepsilon}_{_R}}
\ln(2\sqrt{e}\gamma\varepsilon_{_R})\cr
\approx&
{\textstyle{3\over4}}{\mu_{_p}\eta\over
\tilde{\varepsilon}^2_{_R}}\,{cR_g\over z^2}
\ln(2\sqrt{e}\gamma\Gamma\tilde{\varepsilon}_{_R}).\cr}
\eqno(3.14)$$
Unlike (3.12) in the Thomson limit, the energy loss rate increases
only slowly with increasing $\Gamma$ and $\gamma$. In $K$, the 
characteristic time for the energy loss is 
$$t_{_{\rm KN}}\approx0.03\,{\rm s}\,
\Bigl({0.1\over \eta}\Bigr)\biggl(\!{
\tilde{\varepsilon}_{_R}\over 6\times10^{-5}}\!\biggr)
\Bigl(\!{z\over 10R_g}\!\Bigr)^2\!{\Gamma\gamma\tilde{\varepsilon}_{_R}
\over\ln(3.3\Gamma\gamma\tilde{\varepsilon}_{_R})},\eqno(3.15)$$
where $\Gamma\gamma\tilde{\varepsilon}_{_R}\geq1$.  

Plasmas in the jet are likely to be magnetized. Then, particles
would also radiate synchrotron emission. The energy loss rate 
due to synchrotron emission is given by
$${d\gamma\over dt}\approx{1\over4\pi}{\sigma_{_{\rm T}}
B^2\gamma^2\over m_ec},\eqno(3.16)$$
where $B$ is the magnetic field in the jet frame. Thus, the synchrotron
time scale is
$$t_s\approx3\times10^3\,{\rm s}\Bigl({10\,{\rm G}\over B}\Bigr)^2
\Bigl({10^2\over\gamma}\Bigr).\eqno(3.17)$$
The energy loss due to synchrotron emission can be comparable to
(3.13) and (3.15) only for strong magnetic fields or large $\gamma$.

Because of the short time scales indicated by (3.13) and (3.15) near 
the black hole, for plasmas in the cell to be highly relativistic 
one needs an efficient injection of electron positron pairs, e.g.
through cascades or acceleration. The relevant processes, either 
pair cascades or direct acceleration of electrons (positrons), must
be fast enough to overcome energy loss due to inverse Compton
scattering or synchrotron radiation. Although details of 
acceleration near a black hole are not well understood, the existence
of a rapid acceleration mechanism that is able to accelerate particles to 
ultrarelativistic energies within a short time scale cannot be ruled out.
One possibility is acceleration by a rotation-induced potential drop,
similar to that occuring in pulsars, where particles extracted from the 
neutron star are accelerated to ultrarelativistic energies within a much
shorter time scale than the rotation period (e.g. Michel 1987).  
Particles injected at radial distance $R_{\rm inj}$ to the black hole 
are acclerated to energy $\gamma\approx\sigma^{2/3}\Delta R/R_{\rm inj}$ 
after travelling a distance $\Delta R$, where $\sigma=e\Delta\Phi/m_ec^2$, 
$\Delta\Phi$ is the potential drop which can be estimated from the 
total power output, $L_0$, of the AGN, and $\Delta R/R_{\rm inj}\ll1$ 
(e.g. Michel 1987). Then, the acceleration time $t_{\rm acc}=\Delta R/c$ 
is estimated to be 
$$t_{\rm acc}={\gamma R_{\rm inj}\over c\sigma^{2/3}}\approx 10^{-3}\,{\rm s}
\Bigl({\gamma\over10^4}\Bigr)\Bigl({R_{\rm inj}\over 2R_g}\Bigr),
\eqno(3.18)$$  
where $\sigma^{2/3}\approx 10^{10}$ for 
$L_0=10^{46}\,{\rm erg}\,{\rm s}^{-1}$. Since $t_{\rm acc}<t_{_{\rm IC}}$, 
$t_{_{\rm KN}}$, electrons (positrons) can be accelerated to the 
energies in the KN regime.

For magnetized plasmas, the condition for relativistic plasmas to be 
contained in the jet requires that the gyroradius being less than the
characteristic transverse size, $R_j$, of the jet. Since the gyroradius is
given by $\rho_g=2\times10^7\,{\rm cm}\,(\gamma/10^4)(1\,{\rm G}/B)$ 
where $B$ is the magnetic field, for the characteristic transverse
size of the jet, 
$R_j\approx R_g=1.5\times10^{13}\,{\rm m}\,(M/10^8M_\odot)$, the 
condition $\rho_g< R_j$ should be easily satisfied. It can also be
shown that the approximation made earlier (Eq. 2.5), i.e. neglecting
electron gyration, is valid. The cyclotron time scale $t_c\approx
\rho_g/c\approx10^{-3}\,{\rm s}\,(\gamma/10^4)(1\,{\rm G}/B)$,
is much shorter than $t_{\rm IC}$, $t_{\rm KN}$, $t_s$, and even much 
shorter than the characteristic time of deceleration, which is 
typically $>1\,\rm s$.

\section{Conclusions and discussion}

Radiative deceleration of relativistic jets due to inverse Compton 
scattering is discussed in both the Thomson and KN scattering regimes.
We show that KN scattering is important in deceleration of
ultrarelativistic jets with initial bulk Lorentz $\Gamma>10^3$. 
For $\Gamma\gg1$, near the black hole and in the 
jet frame, the incoming photons are beamed towards the hole, and so 
the main contribution to the drag of the jet is from head-on scattering 
by forward moving particles. These particles are more likely to be in 
the KN regime as they see photons of the highest energy ($\varepsilon'>1$). 
Particles scattering in the Thomson regime may contribute positively to the
radiation force but overall KN scattering is dominant and results in
deceleration. Thus, scattering in the KN regime should be included in 
calculations of the radiation force when $\Gamma>10^3$. Our result shows 
that in the KN regime, radiation drag is reduced, but still severely 
constrains the speed of the jet bulk flow. Thus, Compton drag can
decelerate an $e^\pm$ jet starting with {\it any} $\Gamma>10$ in the region
sufficiently close to the black hole down to the value $\Gamma<10$.
The efficiency of deceleration is significantly reduced if the jet 
contains protons since protons have much smaller scattering cross
section and the effect of protons is to increase the inertia of the jet.

According to the unified scheme (e.g. Urry \& Padovani 1995), 
blazars are radio-loud quasars, and if this is true, a relativistic 
jet with sufficient particle kinetic
luminosity is required to power radio lobes at a larger distance. 
The results given here show that because of radiation drag the 
ability of an electron-positron jet  to power large scale radio 
lobes is severely constrained.  For example, consider an $e^\pm$ jet
with an initial luminosity $\dot{N}_i\Gamma_i \approx L_d$, where 
$\dot{N}_i$ and $\Gamma_i\gg1$ are the initial particle flux and Lorentz
 factor. As a result of Compton drag, the flow slows down to 
$\Gamma_f<10$ with most of the energy going into radiation. Thus, even 
with pair cascades $\dot{N}_f\gg \dot{N}_i$, we may still have 
$\dot{N}_f\Gamma_f\ll \dot{N}_i\Gamma_i$. The constraint can be 
overcome if $e^\pm$ are re-accelerated at a distance further away
from the central region or if there is outflow of accelerated protons 
as suggested in a two-flow model by Sol, Pelletier \& Asseo (1989).
In their model, radio lobes are powered by outflow of p-e plasma jets and
$e^\pm$ jets are dominant only in the subparsec region.

In our discussion, the external soft photons are assumed to come
only from the disc emission which is the sum of blackbody emission from 
series of rings centred at the black hole, and is similar to the 
model discussed by Demer \& Schlickeiser (1993). Generalization of the 
calculation to include other radiation fields, such as that due to a
disc torus (Protheroe \& Biermann 1997), and reprocessed radiation,
should be straightforward. The effect on the jet due to scattering of photons 
from the broad-line region may also be important, and is not considered 
here. In calculating the average force, we assumed an isotropic 
(angular) distribution of electrons (or positrons) with a power law. 
Some mechanism, e.g. pitch angle scattering, is required for 
isotropization since inverse Compton scattering itself tends to make 
the distribution anisotropic.

\section*{Acknowledgements}

QL acknowledges the receipt of Australian Research Council
(ARC) Postdoctoral Fellowship.

\section*{Appendix A: Momemtum transfer due to Compton scattering}

Calculation of momemtum transfer (2.5) requires the full KN differential
cross $d\sigma/d\varepsilon_sd\Omega_s$ which was discussed in details
by Blumenthal \& Gould (1970). Here, we use the following approximation
(in $K_j$) (Reynolds 1982; Dermer \& Schlickeiser 1993)
$${d\sigma\over d\varepsilon_sd\Omega_s}\approx
\sigma_{_{\rm T}}\delta\bigl[\varepsilon_s-\gamma^2\varepsilon(1-
 \beta\cos\Theta)\bigr]\,\delta(\Omega_s-\Omega_e),\eqno({\rm A}1)$$
for the Thomson regime ($\varepsilon'<1$) and
$${d\sigma\over d\varepsilon_sd\Omega_s}\approx
{3\sigma_{_{\rm T}}\over
8\varepsilon'}\ln(2\sqrt{e}\varepsilon')
\delta(\varepsilon_s-\gamma)\delta(\Omega_s-
\Omega_e),\eqno({\rm A}2)$$
for the KN regime ($\varepsilon'\geq1$),
where $\sigma_{_{\rm T}}$ is the Thomson cross section, $\varepsilon'$ 
is the incoming photon energy in the electron rest frame,
$\cos\Theta=\cos\theta\cos\theta_e+\sin\theta\sin\theta_e
\cos(\phi-\phi_e)$, $\Theta$ is the angle between the incoming
photon direction and the electron motion, the delta function 
$\delta(\Omega_s-\Omega_e)$ describes the beaming approximation.

To evaluate (2.5), particles with $\gamma$ can be broadly separated 
into two groups: one group of particles with $\cos\theta_e<\mu_c$) 
scatter in the Thomson regime with (A1) and the other group with 
$\cos\theta_e\geq\mu_c$ scatter in the KN regime with (A2). This 
approximation is similar to that was used Dermer \& Schlickeiser 
(1993) in calculating scattered photon spectrum. Thus, using (A1) 
and (A2), the radiation force (2.5) can be written approximately 
into the form
$$\eqalign{f_z&\equiv\int\!d\Omega_ed\gamma\,{dp_z\over dt}
n_e(\gamma,\Omega_e)\cr
\approx& -\sigma_{_{\rm T}}
\!\int\!dR\,{RF(R)\over2\pi(R^2+z^2)}\cr
\times&
\tilde{\varepsilon}_{_R}\Gamma^2
      (1-\beta_b\cos\tilde{\theta}_{_R})^2
\int\!d\Omega_ed\gamma\,n_e(\gamma,\Omega_e) \cr
\times&\biggl\{\!(1\!-\!\beta\cos\Theta_{_R})\Bigl[
\!\gamma^2(1\!-\!\beta\cos\Theta)\cos\theta_e-\cos\theta_{_R}\Bigr]
   H(1-\varepsilon'_{_R})\cr
+& {\textstyle{3\over8}}{\ln(2\sqrt{e}\varepsilon'_{_R})
\over\Gamma\gamma\tilde{\varepsilon}_{_R}}\bigl(\gamma\cos\theta_e-
\varepsilon_{_R}\cos\theta_{_R})H(\varepsilon'_{_R}-1)\!\biggr\},
\qquad\qquad\llap{({\rm A}3)}\cr}
$$
where $H(\varepsilon'_{_R}-1)=1$ for $\varepsilon'_{_R}\geq1$ and 
$H(\varepsilon'_{_R}-1)=0$ for $\varepsilon'_{_R}<1$, 
$n_e(\gamma,\Omega_e)$ is the particle distribution as defined 
earlier (Eq. 2.7). The terms with $H(1-\varepsilon'_{_R})$ and 
$H(\varepsilon'_{_R}-1)$ correspond respectively to  scattering in 
the Thomson and KN regimes.

Accordingly, $f_z$ is written as a sum of two components,
$f_z=f_{z,{\rm T}}+f_{z,{\rm KN}}$ where $f_{z,{\rm T}}$, 
$f_{z,{\rm KN}}$ describe the contributions to the force from 
scattering in the Thomson and KN regime, respectively, and they are
$$\eqalign{f_{z,{\rm T}}&= -\sigma_{_{\rm T}}
\!\int\!dR\,{RF(R)\over2\pi(R^2+z^2)}
\tilde{\varepsilon}_{_R}\Gamma^2
      (1-\beta_b\cos\tilde{\theta}_{_R})^2 \cr
\times&\int\!d\gamma\int\!{d\phi_e\over2\pi}\int^{\mu_c}_{-1}\!\!
{d\cos\theta_e\over2}\,n_e(\gamma,\Omega_e)
(1\!-\!\beta\cos\Theta_{_R})\cr
\times&\Bigl[
\!\gamma^2(1\!-\!\beta\cos\Theta)\cos\theta_e-\cos\theta_{_R}\Bigr]
,\cr}
\eqno({\rm A}4)$$
and
$$\eqalign{f_{z,{\rm KN}}&= -\sigma_{_{\rm T}}
\!\int\!dR\,{RF(R)\over2\pi(R^2+z^2)}
\tilde{\varepsilon}_{_R}\Gamma^2
      (1-\beta_b\cos\tilde{\theta}_{_R})^2 \cr
\times&
\int\!d\gamma\int\!{d\phi_e\over2\pi}\int^1_{\mu_c}\!\!
{d\cos\theta_e\over2}\,n_e(\gamma,\Omega_e)
{\textstyle{3\over8}}{\ln(2\sqrt{e}\varepsilon'_{_R})
\over\Gamma\gamma\tilde{\varepsilon}_{_R}}\cr
\times&\bigl(\gamma\cos\theta_e-\varepsilon_{_R}\cos\theta_{_R}),
\cr}
\eqno({\rm A}5)$$
where $\mu_c$, $n_e(\gamma,\Omega_e)$ are defined by (2.12), (2.7). 
Integration over $\cos\theta_e$ can be easily done using approximation 
$|\sin\theta_{_R}|\ll|\cos\theta_{_R}|\approx1$, 
$\cos\theta_{_R}\approx-1$. In this approximation, the integrands in 
(A4) and (A5) can be regarded as independent of $\phi_e$. Then, 
integrating over $\cos\theta_e$, we obtain 
$$\eqalign{f_{z,{\rm T}}&=\cr
-&{\textstyle{1\over2}}\sigma_{_{\rm T}}
C_0n_{e0}\!\!
\int\!dR\,{RF\tilde{\varepsilon}_{_R}\Gamma^2\over
2\pi(R^2+z^2)}(1-\beta_b\cos\tilde{\theta}_{_R})^2
\!\!\int\!d\gamma\,\gamma^{-p}\cr
\times&\Bigl\{\gamma^2\Bigl[
{\textstyle{1\over2}}(\mu^2_c-1)-{\textstyle{2\over3}}\beta
\cos\theta_{_R}(\mu^3_c+1)
+{\textstyle{1\over4}}(\mu^4_c-1)\Bigr]\cr
-&\cos\theta_{_R}(\mu_c+1)+
{\textstyle{1\over2}}(\mu^2_c-1)\Bigr\},
\qquad\qquad\qquad\llap{({\rm A}6)}\cr}
$$
$$\eqalign{f_{z,{\rm KN}}&\approx-{\textstyle{3\over8}}\sigma_{_{\rm T}}
C_0n_{e0}
\!\int\!dR\,{RF\over2\pi(R^2+z^2)}{1\over4\tilde{\varepsilon}_{_R}}
\int\!d\gamma\,\gamma^{-p}\cr
&\times\Bigl\{ 
\bigl(1-\mu^2_c\bigr)\ln2-{1-\mu_c\over\beta\cos\theta_{_R}}
\cr
&-{2\varepsilon_{_R}\cos\theta_{_R}\over\gamma}
\Bigl[(1-\mu_c)\ln(2/\sqrt{e})\cr
&+ 
\Bigl(1-{1\over\beta\cos\theta_{_R}}\Bigr)
\ln(2\varepsilon_{_R}\gamma)\Bigr]\Bigr\}
H(\gamma\varepsilon_{_R}-{\textstyle{1\over2}}).\cr}
\eqno({\rm A}7)$$
Eq. (A6) reduces to Eq. (2.6) or (2.8) by assuming $\mu_c=1$,
i.e. all particles are in the Thomson scattering regime.
Let $I_{\rm T}$ and $I_{\rm KN}$ represent $\gamma$-integration in
(A6) and (A7); $I_{\rm T}$ and $I_{\rm KN}$ are calculated to be
$$\eqalign{I_{\rm T}
=&{1\over 2(p+1)\varepsilon^2_R}\Bigl(1+{1\over2\varepsilon^2_R}\Bigr)
\Bigl(\gamma^{-p-1}_c-\gamma^{-p-1}_{_{\rm max}}
\Bigr)\cr
+& {1\over 3p\varepsilon^3_R}\Bigl(\gamma^{-p}_{_{\rm max}}-
\gamma^{-p}_c\Bigr)+{4\over3(3-p)}
\Bigl(\gamma^{3-p}_c-\gamma^{3-p}_{_{\rm min}}\Bigr)\cr
+&{2\over p-1}\Bigl(\gamma^{1-p}_{_{\rm min}}-\gamma^{1-p}_c\Bigr),
\cr}\eqno({\rm A}8)
$$
$$\eqalign{I_{\rm KN}&=
{2\over p-1}\Bigl(\gamma^{1-p}_c-\gamma^{1-p}_{_{\rm max}}\Bigr)\cr
+&{1\over p}\Bigl[
{\ln(4e)\over\varepsilon_R}+2\varepsilon_R\ln(4/e)+4\varepsilon_R/p\Bigr]
\Bigr(\gamma^{-p}_c-\gamma^{-p}_{_{\rm max}}\Bigr)\cr
-&
{1\over p+1}\Bigl[{\ln2\over\varepsilon^2_R}+\ln(4/e)\Bigr]
\Bigl(\gamma^{-p-1}_c-\gamma^{-p-1}_{_{\rm max}}\Bigr)\cr
+&{4\varepsilon_R\over p}\Bigl[\gamma^{-p}_c\ln(2\varepsilon_R\gamma_c)-
\gamma^{-p}_{_{\rm max}}\ln(2\varepsilon_R\gamma_{_{\rm max}})\Bigr],\cr}
\eqno({\rm A}9)$$
where $\gamma_c={\rm max}\{1/\varepsilon_{_R},\gamma_{_{\rm min}}\}$.
Then, (A6) and (A7) can be further written into the form
$$\eqalign{&f_{z,{\rm T}}=\cr
&\!-{\textstyle{1\over2}}\sigma_{_{\rm T}}C_0n_{e0}\!\!
\!\int^{R_{\rm max}}_{R_{\rm min}}\!\!\!
\!dR\,{RF\tilde{\varepsilon}_{_R}\Gamma^2
(1\!-\!\beta_b\cos\tilde{\theta}_{_R})^2\over 2\pi(R^2+z^2)}
I_{\rm T},\cr}\eqno({\rm A}10)$$
$$f_{z,{\rm KN}}\approx-{\textstyle{3\over8}}\sigma_{_{\rm T}}
C_0n_{e0}\!\!\int^{R_{\rm max}}_{R_{\rm min}}\!dR\,{RF\,I_{\rm KN}
\over2\pi(R^2+z^2)}{1\over4\tilde{\varepsilon}_{_R}}.
\eqno({\rm A}11)$$

\vfil
\bye